\PassOptionsToPackage{svgnames}{xcolor}
\documentclass[sigconf]{acmart} 

\usepackage{booktabs} % For formal tables
\usepackage{etoc} % for table of content for appendix only
\usepackage{wrapfig}
\usepackage{amsmath,bbm}
\usepackage{amsfonts}
\usepackage{amssymb}

\usepackage{verbatim}
\usepackage{graphicx}
\usepackage{subfig}
\usepackage{epstopdf}
\DeclareGraphicsExtensions{.pdf,.eps,.png,.jpg,.tif,.tiff,.ps}
\graphicspath{{img//}}

\usepackage{footnote}
\makesavenoteenv{tabular}
\makesavenoteenv{table}

\usepackage{soul}
\usepackage[colorinlistoftodos]{todonotes}
\usepackage[svgnames]{xcolor}

\usepackage{xspace}

\newcommand{\hipdemo}{HIPie\xspace}

\definecolor{navy}{rgb}{0.1, 0.1, 0.8}
\definecolor{ruby}{rgb}{0.8, 0.2, 0.2}
\definecolor{gray}{rgb}{0.4, 0.4, 0.4}
\definecolor{myblue}{rgb}{.8, .8, 1}
\definecolor{olive}{rgb}{0.1, 0.5, 0.1}

\newcommand{\eat}[1]{}
\newcommand{\rev}[1]{{#1}}

\newcommand{\verify}[1]{{}}
\newcommand{\NOTE}[2]{ }
\newcommand{\TODO}[2]{}
\newcommand{\nb}[1]{}

\newcommand{\secmoveup}{\vspace{-0mm}} % {\vspace{-1.5mm}}
\newcommand{\eqmoveup}{\vspace{-0mm}} %{\vspace{-2.4mm}}
\newcommand{\captionmoveup}{\eqmoveup\vspace{-0mm}}   %{\vspace{-2.4mm}}

\newcommand{\squishlist}{
 \begin{list}{$\bullet$}
  { \setlength{\itemsep}{0pt}
     \setlength{\parsep}{3pt}
     \setlength{\topsep}{3pt}
     \setlength{\partopsep}{0pt}
     \setlength{\leftmargin}{1.5em}
     \setlength{\labelwidth}{1em}
     \setlength{\labelsep}{0.5em} } }

\newcommand{\squishlisttwo}{
 \begin{list}{$\bullet$}
  { \setlength{\itemsep}{0pt}
    \setlength{\parsep}{0pt}
    \setlength{\topsep}{0pt}
    \setlength{\partopsep}{0pt}
    \setlength{\leftmargin}{1.5em}
    \setlength{\labelwidth}{1.5em}
    \setlength{\labelsep}{0.5em} } }

\newcommand{\squishend}{
  \end{list}  }

\copyrightyear{2018}
\acmYear{2018}
\setcopyright{iw3c2w3}
\acmConference[WWW '18 Companion]{The 2018 Web Conference Companion}{April
23--27, 2018}{Lyon, France}
\acmBooktitle{WWW '18 Companion: The 2018 Web Conference Companion, April
23--27, 2018, Lyon, France}
\acmPrice{}
\acmDOI{10.1145/3184558.3186972}
\acmISBN{978-1-4503-5640-4/18/04}

\begin{document}
%% just a tag, to let know the etoc package not to show these sections.
\etocdepthtag.toc{mtchapter} 

%% MAR: title versions
\title{\hipdemo: Hawkes Intensity Process Visualizer}
\title{\hipdemo: Explaining and Predicting the Virality of Youtube Videos}
\title{\hipdemo: Enabling Reasoning about Youtube Video Virality}
\title{Will This Video Go Viral? Explaining and Predicting the Popularity of Youtube Videos}

%\titlenote{Produces the permission block, and
%  copyright information}
%\subtitle{Extended Abstract}
%\subtitlenote{The full version of the author's guide is available as
%  \texttt{acmart.pdf} document}

%\author{Quyu Kong \hspace{1cm} Marian-Andrei Rizoiu \hspace{1cm} Siqi Wu \hspace{1cm} Lexing Xie \\
%ANU \& Data61 CSIRO, Canberra, Australia.}

\author{Quyu Kong}
\affiliation{%
  \institution{ANU \& Data61 CSIRO}
 \city{Canberra}
 \country{Australia}
}

\author{Marian-Andrei Rizoiu}
\orcid{orcid.org/0000-0003-0381-669X}
\affiliation{%
  \institution{ANU \& Data61 CSIRO}
 \city{Canberra}
 \country{Australia}
}

\author{Siqi Wu}
\affiliation{%
  \institution{ANU \& Data61 CSIRO}
 \city{Canberra}
 \country{Australia}
}

\author{Lexing Xie}
\affiliation{%
  \institution{ANU \& Data61 CSIRO}
 \city{Canberra}
 \country{Australia}
}

%\renewcommand{\shortauthors}{Kong et al.}
%\renewcommand{\shorttitle}{Explaining and Predicting the Popularity of Youtube Videos}

\begin{abstract}
	%!TEX root = hip-demo.tex
%

What makes content go viral? 
Which videos become popular and why others don't?
Such questions have elicited significant attention from both researchers and industry, particularly in the context of online media.
A range of models have been recently proposed to explain and predict popularity;
however, there is a short supply of practical tools, accessible for regular users, that leverage these theoretical results.
\textbf{\hipdemo} -- an interactive visualization system -- is created to fill this gap, by enabling users to reason about the virality and the popularity of online videos.
It retrieves the metadata and the past popularity series of Youtube videos, it employs \rev{the} Hawkes Intensity Process, a state-of-the-art online popularity model for explaining and predicting video popularity, and it presents videos comparatively in a series of interactive plots.
This system will help
both content consumers and content producers in a range of data-driven inquiries, such as to comparatively analyze videos and channels, to explain and \rev{to} predict future popularity, to identify viral videos, and to estimate \rev{responses} to online promotion.

\end{abstract}

%\begin{CCSXML}
%<ccs2012>
%  <concept_id>10010520.10010553.10010562</concept_id>
%  <concept_desc>Computer systems organization~Embedded systems</concept_desc>
%  <concept_significance>500</concept_significance>
%  <concept_id>10010520.10010575.10010755</concept_id>
%  <concept_desc>Computer systems organization~Redundancy</concept_desc>
%  <concept_significance>300</concept_significance>
%  <concept_id>10010520.10010553.10010554</concept_id>
%  <concept_desc>Computer systems organization~Robotics</concept_desc>
%  <concept_significance>100</concept_significance>
%  <concept_id>10003033.10003083.10003095</concept_id>
%  <concept_desc>Networks~Network reliability</concept_desc>
%  <concept_significance>100</concept_significance>
%</ccs2012>  
%\end{CCSXML}

%% MAR: don't think they're needed for submission
%\ccsdesc[500]{Computer systems organization~Embedded systems}
%\keywords{ACM proceedings, \LaTeX, text tagging}

\maketitle

%\input{samplebody-conf}

%\begin{figure*}[tbp]
%	\centering
%	\includegraphics[width=0.9\textwidth]{phi-SIR-Hawkes}
%	
%	\caption{An example rate of new events/new infections for the Hawkes model (left) and SIR model (right), generated by a single event/individual having occurred at time $t_i = 0$.}
%	\label{fig:example-phi-Hawkes-SIR}
%\end{figure*}

%!TEX root = hip-demo.tex

\secmoveup
\section{Introduction}

The popularity of online videos is typically measured by the number of views they attract from viewers.
Understanding online popularity can help content producers to propose better content, and content consumers to deal with information overload.
Viral videos quickly catch the attention of the viewers and achieve very high popularity in short periods of time.
Explaining what makes videos go viral, and identifying them early would prove useful for advertisers and content providers.

%Popularity refers to the attraction (e.g. number of views) drawn by online cultural items, such as videos and blog posts. Modeling popularity can help online content providers measure the content quality of online items and online content consumers deal with information overload. A range of models have been proposed for modeling online item popularity including the Hawkes Intensity Process (HIP)~\cite{Rizoiu2017}.

% \TODO{MAR}{Why did we build the demo? What open questions are there?\\
% 1. Explain popularity over time for online content; and forecast popularity;\\
% 2. for consumers: Compare videos, selected on the fly from Youtube;\\
% 3. for content producers/managers/promoters: Quantify virality and simulate reaction to promotion.}

% \TODO{MAR}{Build a system to visualize online popularity; quantify and compare online virality}

% \TODO{MAR}{This part needs revising, towards how the demo answers the open question.}

Our tool aims to fill several gaps about the systems that enable users to reason about the popularity of online videos.
The first gap concerns the availability of such systems.
Despite the range of theoretical models that have been recently proposed for modeling online popularity~\cite{bakshy2011everyone,Martin2016,Zhao2015,Rizoiu2017},
% popularity including the Hawkes Intensity Process (HIP)~\cite{Rizoiu2017}.
%This work is mainly concerned with addressing three open questions about applying popularity modeling for online contents in real life. 
%Although there are many theoretical models available for modeling online content popularity, 
\textbf{there is no readily available software that allows regular users to easily
%the first question generally still interests every ordinary user:
examine the popularity over time for online videos and forecast their future popularity.}
The second gap concerns content producers and advertisers who need to choose which videos to promote and to identify potentially viral videos.
\textbf{How can content producers quantify virality and simulate video reaction to online promotions?}
The third gap sits for content consumers. 
Most distribution platforms (e.g. Youtube) feature personalized recommendation systems; these usually act as black boxes and make the decision for the user.
%when they are making a decision about interesting online contents. 
%While most current recommendation systems, such as the recommendations in homepage of Youtube, act as a black-box, 
The open question is \textbf{how can the user be empowered by enabling her to compare and select content on the fly?}
%On the other hand, content producers need an efficient way to 

In this work, we answer the above three questions, building upon the current state-of-the-art popularity model, the Hawkes Intensity Process (HIP)~\cite{Rizoiu2017}.
We introduce the \emph{HIP Insights Explorer} (\hipdemo), 
%In this work, we address all three questions by introducing 
an interactive web-based application designed to assist users to reason about the popularity and the virality of Youtube videos.
%for visualizing online popularity, quantifying and comparing online virality. 
%Dealing specifically with Youtube video data, one can use \hipdemo as a tool to address those open questions by extracting information from the visual presentation of videos. 
It exposes a series of measures derived from HIP -- such as the sensitivity to external promotions and the endogenous amplification.\eat{ -- and it}
It allows \rev{one} to
%This tool also allows one to 
conduct various tasks, including identifying prospective popular videos, simulating video reaction to promotion schedules, comparing videos from different authors (i.e. Youtube channels) and visualizing the popularity series fitted and predicted by HIP.
The most important visualization of \hipdemo is the endo-exo map~\citep{Rizoiu2017}, a projection of videos in the two-dimensional space defined by \rev{the} endogenous response and exogenous sensitivity.
The relative positions of the videos in this space indicate their potential of becoming viral.
%The default installation contains 212 Youtube videos already analyzed with HIP.
\hipdemo allows adding any Youtube video on-the-fly as long as its popularity series are available.
%
%but, with the HIP model built into the server side of \hipdemo, \hipdemo allows one to model a new video as long as the historical statistics is available.

The main contributions of this work include:
%\begin{itemize}
\squishlist
	\item A web-based interactive tool to visualize and predict future video popularity using the HIP~\citep{Rizoiu2017} popularity model; 
%	\item A web-based interactive visualization tool for visualizing online content popularity. With the state-of-art popularity model HIP as the core, it allows one to quantify and compare online virality.
	
	\item The endo-exo map visualization, on which \rev{the} viral potential of videos is compared;
	
	\item \hipdemo enables a series of \rev{applications} concerning online popularity, such as comparing videos and channels, identifying future popular videos and simulating video reaction to promotion.

%    \item This work enables content consumers to deal with online content selection by comparing Youtube videos and Youtube channels in an intuitive popularity plot where positions of videos explain popularity potentials.
%
%    \item Content providers can leverage this work on choosing Youtube videos with high popularity potentials to promote. This is achieved in \hipdemo by identifying future popular videos and simulating promotions on videos.
%    \item As an open source project\footnote{Github repository of \hipdemo: https://github.com/computationalmedia/hipie/}, this work presents an example of the construction of a complex web-based visualization platform in R language.
%\end{itemize}
\squishend

%!TEX root = hip-demo.tex

%!TEX root = hip-demo.tex

\begin{figure*}[tbp]
	\centering
	\newcommand\myheight{0.14}
	\newcommand\myheighta{0.162}
	\subfloat[] {
		\includegraphics[height=\myheight\textheight]{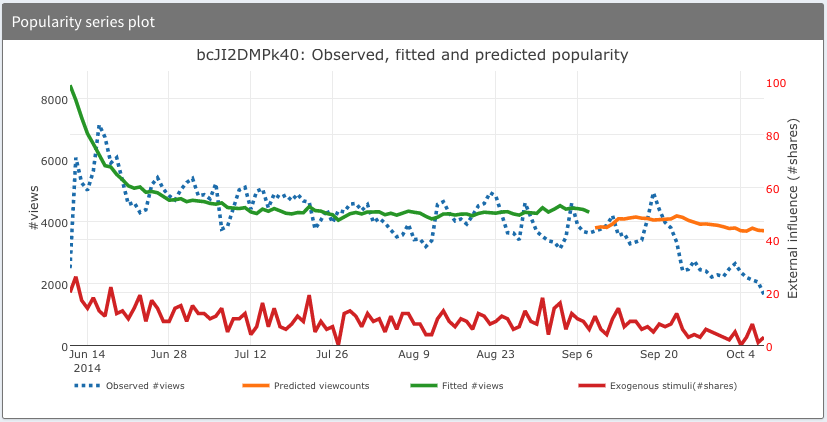}
		\label{subfig:popularity-series}
 	}
	\subfloat[] {
		\includegraphics[height=\myheight\textheight]{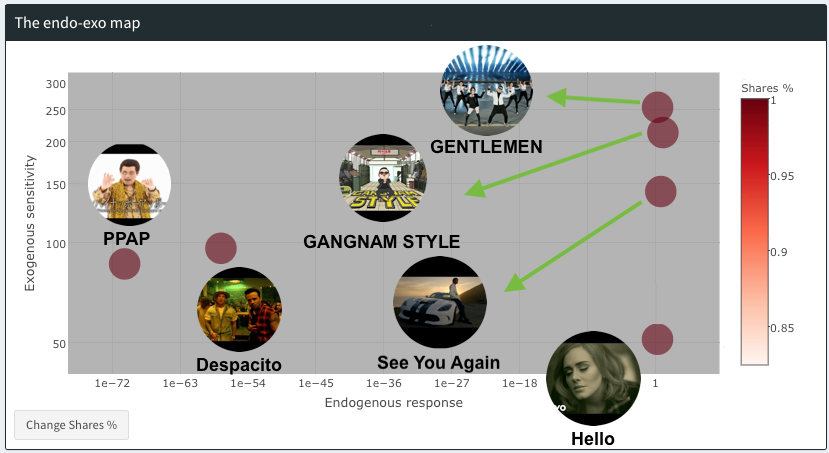}
		\label{subfig:endo-exo}
	} 
	\subfloat[] {
		\includegraphics[height=\myheight\textheight]{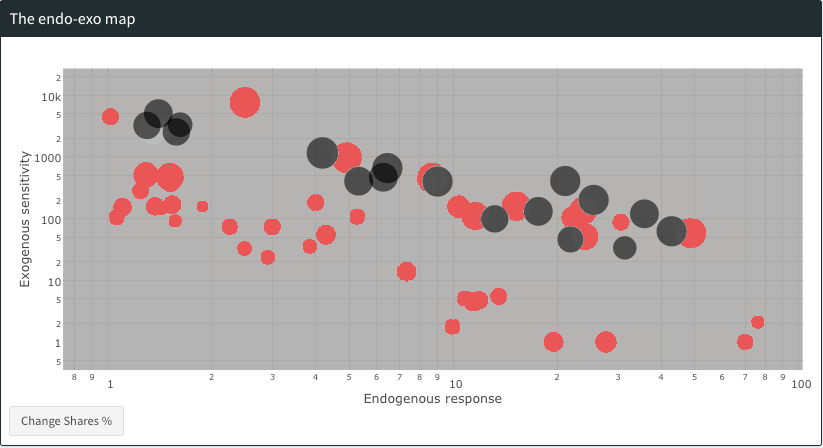}
		\label{subfig:separating-channels}
	}\\
	\subfloat[] {
		\includegraphics[height=\myheighta\textheight]{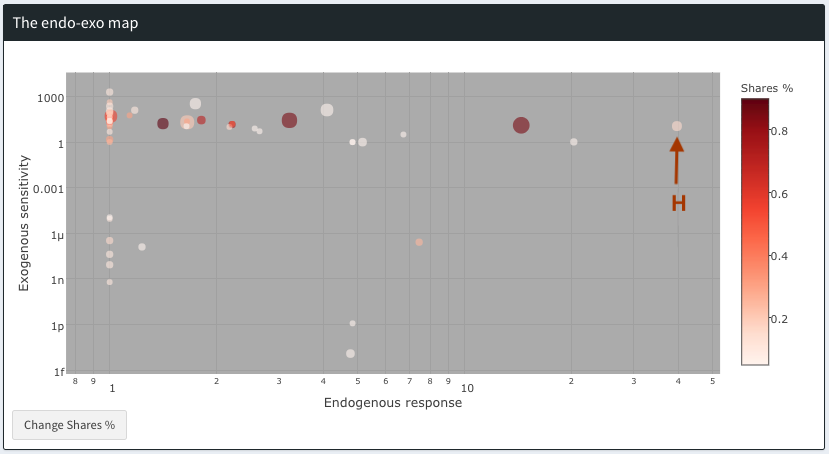}
		\label{subfig:future-popular}
	} 
    \subfloat[] {
		\includegraphics[height=\myheighta\textheight]{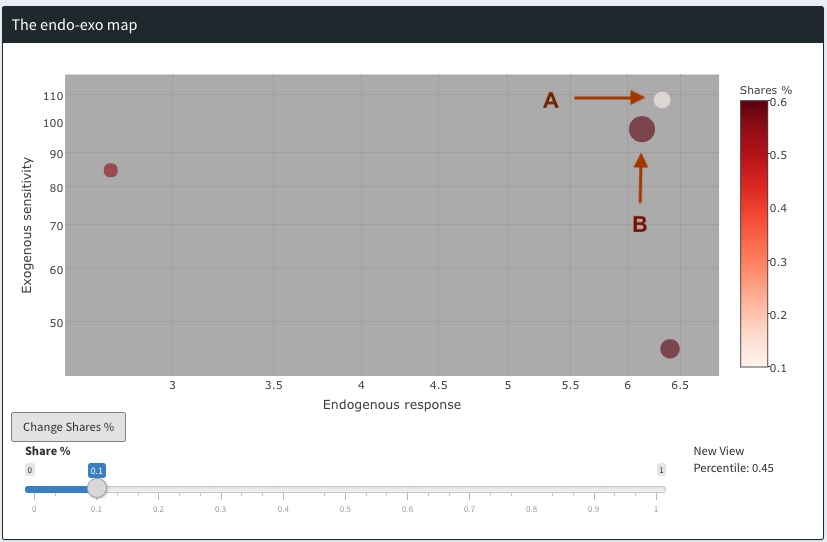}
		\label{subfig:promotions-before}
	}
	\subfloat[] {
		\includegraphics[height=\myheighta\textheight]{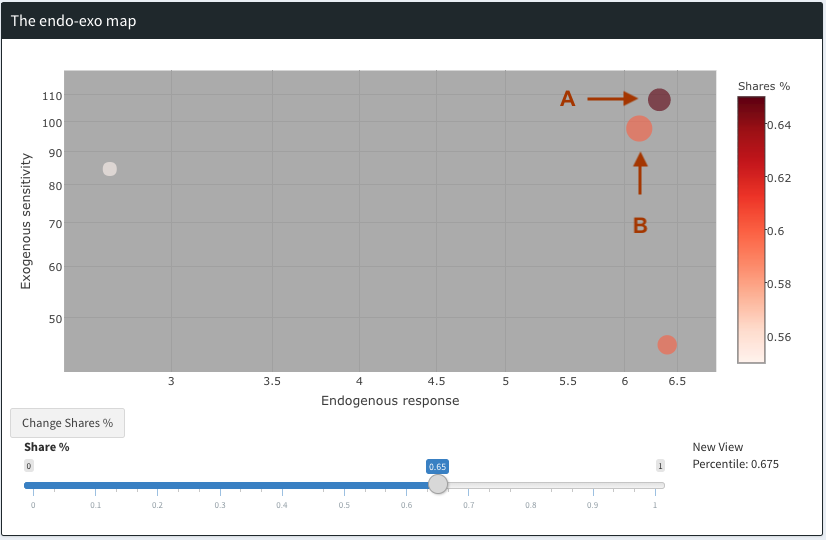}
		\label{subfig:promotions}
	}
	\caption{ 
		Applications enabled by \hipdemo.
		\textbf{(a)} Fitting and forecasting future popularity.
		\textbf{(b)} Using the endo-exo map to explore video virality. 
		\textbf{(c)} Separating the videos of two channels.
        \textbf{(d)} Identifying potentially viral videos.
        \textbf{(e)(f)} Simulating reaction to promotions for video A.
		The size of the bubble shows video popularity before the promotion \textbf{(e)} and after the promotion \textbf{(f)}.
%        \textbf{(f)} Simulating promotions (after simulation).
	}
	\label{fig:sir-expected-observed}
	\captionmoveup
\end{figure*}

\secmoveup
\section{Prerequisites}

In this section we briefly review the HIP model and how it is used to quantify virality and simulate the effect of promotions.

\textbf{Explain and forecast popularity with HIP.}
HIP~\citep{Rizoiu2017} is a novel generative model that explains online popularity series by linking exogenous inputs from public social media platforms, such as Twitter or Youtube, to endogenous responses within the Youtube content platform,
%The endogenous responses 
which account for the word-of-mouth process occurring around videos.
%
%uses a generative model to describe the continuous interaction between external promotions (e.g. tweets or shares about a video) and popularity dynamics (i.e. daily views).
%missing 
%
%\TODO{MAR}{Lexing proposes that we explain HIP without the equation.}
%HIP\citep{Rizoiu2017} is a generative model for explaining the underlying popularity properties of social media items by quantifying the endogenous and exogenous factors hidden in the diffusion process of those items. 
%It is obtained by taking expectation over stochastic event histories and converting marked Hawkes self-exciting processes to a model that describes expected event volumes over time. It is expressed as the following 
%At the heart of HIP lies the following equation:
HIP models popularity using:
\begin{equation} \label{eq:HIP}
	\eqmoveup
	\xi(t) = \mu s(t) + C \int^{t}_{0} \xi(t-\tau)(\tau + c)^{-(1+\theta)}d\tau
\end{equation}
where $\xi(t)$ is the number of views that the video receives during day $t$ and $s(t)$ is the volume of exogenous inputs (tweets, shares or promotions).
Eq.~\eqref{eq:HIP} can intuitively be understood as: the number of views a video receives during day $t$ is dependent on its popularity at each previous day $1, .. , t-1$ decayed by how fast people forget ($\theta$ is the exponent controlling the power-law decay of social memory).
The parameters of HIP ($\mu, C, c, \theta$) are fit on an observed prefix of the views and the shares series.
If future exogenous stimuli \rev{are} known, the popularity series can be ``run forward'' by plugging $s(t)$ and the past popularity into Eq.~\eqref{eq:HIP}.
%HIP can be also applied to forecast future view trending of a video given its history data. \TODO{MAR}{How?}

\textbf{The endo-exo map.}
Two metrics derived from HIP describe a video's virality. 
%- the endogenous response $\mu$ and exogenous sensitivity $A_{\hat{\xi}}$ - which are helpful to identify items with high popularity potential \cite{Rizoiu2017}. 
The exogenous sensitivity $\mu$ quantifies the video sensitivity to the external stimuli $s(t)$.
The endogenous response $A_{\hat{\xi}}$ is computed as $A_{\hat{\xi}} = \sum^{\infty}_{t=0} \hat{\xi}(t)$, where $\hat{\xi}(t)$ is the popularity series generated by a single initial exogenous impulse.
Intuitively, $A_{\hat{\xi}}$ represents the total amount of endogenous amplification that each view generates.
%defines a video's reaction to external stimulation. \TODO{MAR}{Intuition: views per promotion!}
\citet{Rizoiu2017} introduce the \emph{endo-exo map}, which is a two-dimensional space of the exogenous sensitivity and the endogenous response. 
It is used to identify potentially viral videos, videos with high scores on both dimensions, as well as unpromotable videos.
%videos that are not promotable.
%\TODO{MAR}{Describe endo-exo map, how do we use it to compare videos?}

%\textbf{Virality and video reaction to promotions.}
\textbf{Viral potential and the reaction to promotions.}
\citet{Rizoiu2017b} use HIP in an advertisement application, in which the aim is to quantify the effect of promotion on content popularity.
The \emph{viral potential} $\nu = \mu A_{\hat\xi}$ is the \emph{return on investment}, or the total number of views generated by a single promotion.
They also study the effect of promotion schedules on the views series.
They construct the promotion series by allocating \rev{to} each day an amount of promotions, and by \rev{introducing} it into HIP alongside with the organic exogenous stimuli to obtain the promoted view series.
%They study the impact of the promotion schedule on the total popularity.

%!TEX root = hip-demo.tex

\section{Applications}
\label{sec:applications}

\hipdemo has a series of functionalities that enable users to understand, reason and interact with the popularity of Youtube videos. 
%These applications fill in the gap of addressing the open questions we bring up earlier.

\textbf{Explain and predict Youtube video popularity (Fig.~\ref{subfig:popularity-series}).} 
For any video, \hipdemo depicts several popularity series: observed, fitted and forecasted by HIP.
% model explains it and what is the predicted future popularity in time series. 
Fig.~\ref{subfig:popularity-series} shows the example of a Music video ``Footprints'' from the Dutch DJ \textit{Tiesto} (Youtube id: \textit{bcJI2DMPk40}).
It shows the popularity series for the first 120 days after video upload.
The dotted blue line represents the observed view counts (i.e., real data) and the red line is the external promotion series.
Fitting HIP and forecasting future popularity are performed in a temporal holdout setup.
The views and shares series in the first 90 days are used to fit the parameters of HIP.
The green line shows the fitted view count series.
The orange line represents the predicted view counts series, using the previously fitted parameters and the external promotion series from day 91 to 120.
\rev{As shown in Fig.~{\ref{subfig:popularity-series}} and in the online public installation (described in Sec.~{\ref{sc:obtaining}}), the popularity series fitted by HIP follows closely the observed popularity series.
Furthermore, \citet{Rizoiu2017} have shown HIP to be able to forecast future popularity with less than $5\%$ mean absolute percentile error when using shares as the exogenous stimuli series, and $5.35\%$ when using tweets.}
%To generate the popularity series, HIP uses the views and shares series in the first 90 days for modeling and uses the shares in the next 30 days to forecast views. 

\textbf{Compare videos (Fig.~\ref{subfig:endo-exo}).} 
\hipdemo enables users to comparatively analyze videos using the endo-exo map, by showing the amount of views and shares they receive, alongside with the exogenous sensitivity and the endogenous reaction.
%The endo-exo map in \hipdemo 
%provides an intuitive explanation for video popularity by representing videos with
%multiple dimensions including view counts, share counts, and, more importantly, endo-exo values. 
Fig.~\ref{subfig:endo-exo} shows six of the most popular \rev{pop} songs on Youtube on the endo-exo map.
Each video is presented as a bubble, where the x coordinate is the endogenous response and the y coordinate is the exogenous sensitivity. 
The color depth of bubbles indicates the amount (in percentage scale) of external promotions that the video receives and the size of bubbles shows the amount of views it receives.
``Gentleman'' (id \textit{ASO\_zypdnsQ}) and ``Gangnam style'' (id \textit{	9bZkp7q19f0}) from the Korean singer Psy occupy the most privileged position on the map, both having a high exogenous sensitivity and a high endogenous response.
In comparison, ``Hello'' by Adele (id \textit{YQHsXMglC9A}) has lower exogenous sensitivity, while ``PPAP'' by the Japanese singer Pikotaro (id \textit{0E00Zuayv9Q}) has a lower endogenous response.
%is an example of comparing some viral music videos from past few years, such as ``PPAP'' and 
%Comparison of these videos shows that the popularity potential differs among the videos. ``GENTLEMAN'' appears to be the video that is most responsive to exogenous stimuli. All videos are generally divided into two groups: ``PPAP'' and ``Despacito'' are the 2 with low endogenous response values but high exogenous sensitivity values, while other videos have both high values.

\textbf{Compare channels (Fig.~\ref{subfig:separating-channels}).} 
In \hipdemo we use the endo-exo map to visualize groups of videos that belong to the same user-assigned content type, or are from the same author (called channel in YouTube). 
Fig.~\ref{subfig:separating-channels} shows in black color a scatter plot of videos in the category News\&Activism, posted by the reporter \emph{Anatolii Sharij} covering the 2014 events in Ukraine, and in red color a user (\emph{VEGETTA777}) focusing on recordings of Game sessions. 
The game recording videos are generally more popular (bigger bubbles) than the news
videos, and this is explained by the former group having higher exogenous sensitivity -- higher values of $\mu$.
%In Youtube, a ``channel'' is basically a denomination for authors/uploaders. 
%One of our default datasets shows that videos from different channels are likely to have different display properties in endo-exo map. 
%Therefore, one can use endo-exo map to separate videos from different channels. 
%For instance, in Fig.~\ref{subfig:separating-channels}, videos from a reporter covering events in Ukraine (\emph{Anatolii Sharij}, in black color) are separated from Spanish game recording videos channel (\emph{VEGETTA777}, in red color) where \emph{VEGETTA777} videos are shown  more popular than \emph{Anatolii Sharij} videos which is explained by \emph{VEGETTA777} videos having higher exogenous sensitivity values.

\textbf{Identify potentially viral videos (Fig.~\ref{subfig:future-popular}).} 
\hipdemo allows to identity videos that have the potential of going viral, but are yet to.
These are videos with high exogenous sensitivity and high endogenous response (top-right corner of the endo-exo map), but which \rev{have received} very little external stimuli.
Video \emph{H} from in Fig.~\ref{subfig:future-popular} -- a Japaneese Film\&Animation video -- is an example of a potentially viral video: it has both high endogenous response and exogenous sensitivity, it has received few promotions (light color) and has achieved little popularity (small bubble size).
Every external stimuli that this video receives will generates \eat{in time }450 views.
%
%As discussed, endogenous response and exogenous sensitivity values together indicate the potential of a new video going viral. 
%For this reason, one is able to identify prospective popular videos by leveraging the endo-exo map. 
%In general, videos with high popularity potentials locate close to the top-right corner in the map, while videos sit close to the left-bottom corner are less likely to become viral. 
%For example, video \emph{A} from in Fig. \ref{subfig:future-popular} has very high potential to go viral as it has both high endogenous response and exogenous sensitivity values and it has not received much promotions yet.

\textbf{Simulate video response to promotions (Fig.~\ref{subfig:promotions-before} \&~\ref{subfig:promotions}).} 
\hipdemo allows simulating ``what-if'' scenarios: what would be the popularity of a video if it \rev{received} an additional volume of external stimulation.
One can promote (or demote) a video by adding or subtracting a volume of promotion, spread equally among the first 90 days (the \emph{even} promotion schedule studied by \citet{Rizoiu2017b}).
Video \textit{A} in Fig.~\ref{subfig:promotions-before} is a collection of ``ice bucket'' challenges (id: \textit{3hSIh-tbiKE}) which receives little external stimuli (light color) and low popularity (small bubble).
Video \textit{B} (a Gaming video, id \textit{0lTTWeavl1c}) has a similar position on the endo-exo map, but it has a higher popularity due to having received more external stimuli.
After promoting video \textit{A} with an amount of promotions similar to \textit{B} (i.e. similar color in Fig.~\ref{subfig:promotions}), \textit{A} achieves a similar popularity \rev{level} as \textit{B} (similar bubble size).
%\TODO{MAR}{ \textit{Agents of SHIELD Italia} (\textit{Youtube video id}: 3hSIh-tbiKE)}
%\hipdemo gives the possibility to simulate promotions for a given video so that one can conduct more intuitive comparison. 
%When promoting or de-promoting a video, the backend will convert share percentiles into number of shares and allocate these shares evenly into the first 90 days, before retraining the HIP model \cite{Rizoiu2017b}.
%The example from Fig.~\ref{subfig:promotions-before} and Fig.~\ref{subfig:promotions} shows that how will the view count change when the amount of shares video \emph{A} received changes. 
%Before changing promotions for video \emph{A}, it has much smaller view count and share count than video \emph{B} but same popularity potential indicated by the positions of these two videos in the endo-exo map. 
%However, when receiving same amount of promotions, video \emph{A} will actually have almost the same total popularity as video \emph{B} which explains the fact that given sufficient promotions, an unpopular video with high potential can go viral.

%!TEX root = hip-demo.tex

%!TEX root = hip-demo.tex

\begin{figure}[tbp]
	\centering
	\newcommand\myheight{0.20}
    \includegraphics[height=\myheight\textheight]{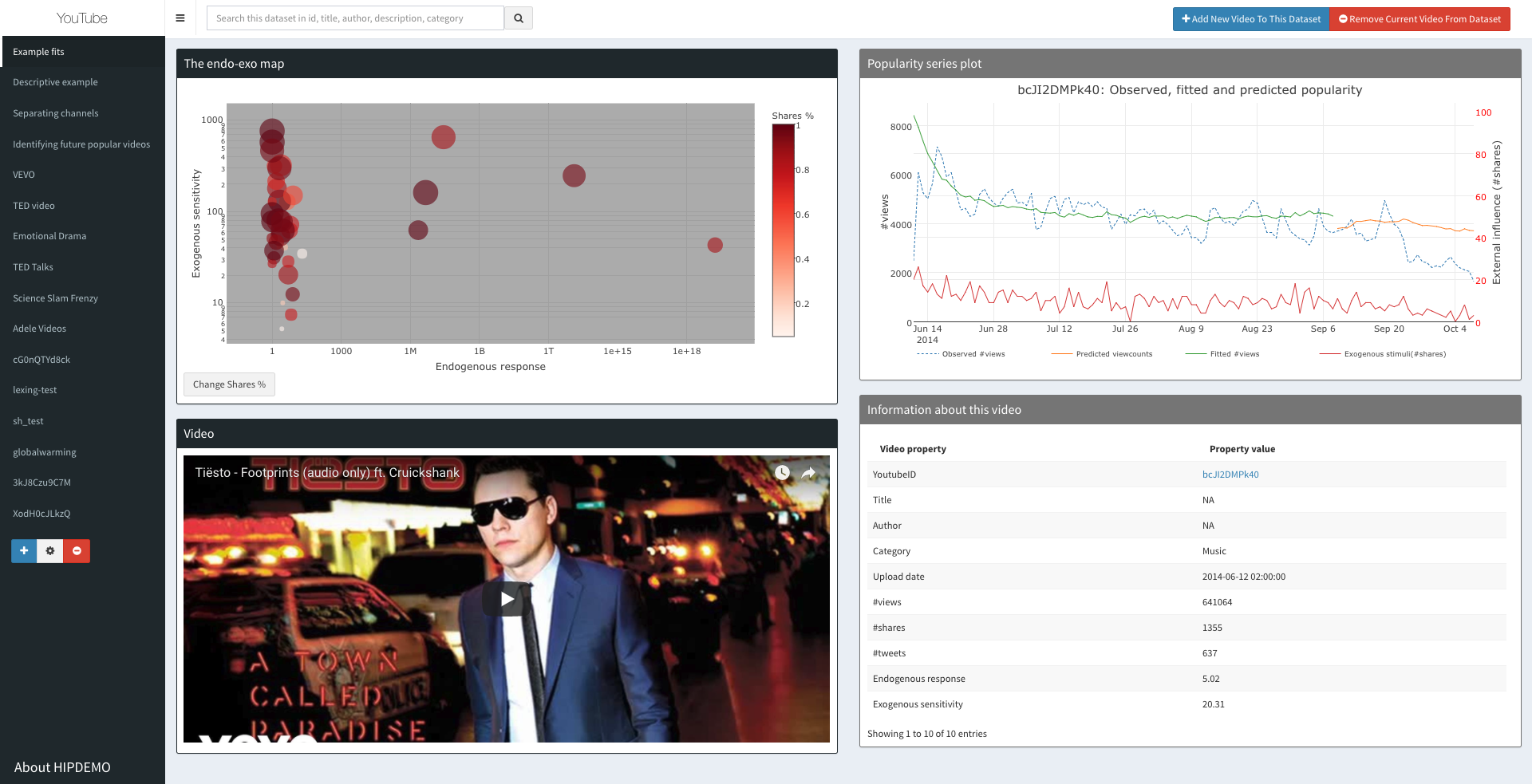}
	\caption{The main interface of \hipdemo.}
	\label{fig:overview}
	\captionmoveup
\end{figure}

\secmoveup 
\section{Description of the demo}

In this section, we introduce the main interfaces of \hipdemo (Sec.~\ref{sc:interfaces}), 
present some implementation choices (Sec.~\ref{sc:implementation}) 
and how to obtain, test and use \hipdemo (Sec.~\ref{sc:obtaining}).

\subsection{Main Interfaces}
\label{sc:interfaces}

\hipdemo is designed as an interactive web application.
%takes the advantage of a web app to provide an interactive experience. 
Fig.~\ref{fig:overview} gives an overview of the main interface of \hipdemo, containing four panels on the right and a navigation menu on the left.
In this section, we describe the main interfaces and functionalities.

% \TODO{MAR}{Most of the info here below is already in Sec.~{\ref{sec:applications}}. Instead of duplicating, say that we use the panels constructed earlier, link and complete information. }

\textbf{The endo-exo map (top-left)}. 
% in Fig.~\ref{fig:overview}, Fig.~\ref{subfig:endo-exo}
This is the most important panel in \hipdemo.
In addition to the characteristics described in Sec.~\ref{sec:applications}, the plot is interactive:
hovering \rev{over} bubbles shows a pop up with additional information about \rev{the} video (YoutubeID, Author, Title, endogenous response and exogenous sensitivity values, acquired percentiles of views and shares).
The user can zoom into certain areas of the map and drag the map.
Clicking on bubbles in endo-exo map causes \rev{other three panels} to switch to the current active video.
The rest of the panels in Fig.~\ref{fig:overview} are updated to show detailed information of the current active video.
%Clicking on a bubble make the corresponding video the current active video, and the other panels (described here below) show its information.
%As other panels are presenting information of a specific video, one can click each bubble in endo-exo map to switch between videos.

\textbf{The popularity panel (top-right)}
%in Fig.~\ref{fig:overview}, Fig.~\ref{subfig:popularity-series}
shows the popularity series of the current active video: observed, fitted and forecasted view counts and observed share counts.
Hovering \rev{over} the plot gives the values of each series at the hovering time point.
% (coordinate on the x-axis). 
%through the plot will give corresponding values at the day for lines.

\textbf{The preview panel (bottom-left).} 
This panel allows to play the selected video, while interacting with the other panels.

\textbf{The video metadata panel (bottom-right).} 
This panel shows detailed metadata information about the current active video: Youtube ID, Title, Author, Category, Upload date, number of views, number of shares and endo-exo values.

\textbf{Collection management menu.} 
\hipdemo allows \rev{one} to create, manage and visualize multiple video collections, using the navigation menu and controls located on the left side of the interface.
%provides the option to switch between different datasets. 
The default collections cannot be changed.
%Apart from the top five default datasets, users can use three buttons to add, modify and delete extra datasets.

\textbf{Adding New Videos.}
\hipdemo allows \rev{one} to add or remove videos from the current collection, by using the buttons in the top-right corner of the interface.
%two buttons sited in the top-right allow one to add or delete a video from a chosen dataset. 
A new video is added by inputing its
% A modal will pop up, after clicking ``Add New Video To This Dataset'' button, requesting a 
Youtube video ID or video link. 
A backend process will crawl the video metadata and popularity series using the \textit{youtube-insight}~\cite{wu2017beyond} package.
Once crawling completes, the first 90 days of the popularity series are used to fit the parameters of HIP. 
Any Youtube videos can be added to the system as long as its popularity series are publicly accessible and at least 120 days data.
After the fitting is completed, the video appears in the corresponding collection.

\subsection{Implementation}
\label{sc:implementation}

\hipdemo is built in R~\citep{R} using the open-source package Shiny~\citep{shiny}, which is dedicated to creating interactive web applications in R.
Shiny enables developers to focus on visualization by simplifying front-end design and backend configuration and it provides web-based input and output tools.
\rev{The other employed packages are chosen for their compatibility with Shiny and for their efficiency}.
\rev{For example, we} employ Plotly to construct the interactive visualization experience. 
\rev{It allows \rev{one} to easily convert static R plots into interactive visualization.}
We also use ShinyJs to build customized JavaScript interactions between R and the frontend, \rev{as it is compatible with default Shiny elements and has many useful APIs}.
Table~\ref{t:packages} lists all the packages employed in \hipdemo, together with their project links and brief descriptions.
%Although Shiny has implemented most frontend frameworks, we still created a few JavaScript functions for custom purposed interaction between R language and frontend codes. For this reason, we included  as well which fills in the gap between JavaScript and R.

\subsection{Obtaining and running the demo}
\label{sc:obtaining}
\hipdemo is open-source and publicly accessible.
To download the source code, access a live demo or a quick tour, we provide the following options:
%It can be downloaded, used and accessed using the following options:
%We provide the following three resources:
\begin{itemize}
	\item A \textbf{public installation} of \hipdemo\footnote{\hipdemo public installation: \url{http://www.hipie.ml/}} is live for testing, requiring only a web browser.
%	This is the easiest way to access the demo, as it only requires a web browser.
    
    \item The source code of \hipdemo can be accessible from a \textbf{Github repository}\footnote{Github repository of \hipdemo: \url{https://github.com/computationalmedia/hipie/}}.
    Simply clone the repository and follow the instructions in the README file to run a local installation.

    \item For a quick tour of the various usages and capabilities of \hipdemo, we provide a \textbf{short Youtube video}\footnote{Screencast for \hipdemo: \url{https://youtu.be/x5xIf4vUScI/}}.
%    those who want to take quick tour through our demonstration, a web accessible video is available on 
\end{itemize}

%!TEX root = hip-demo.tex

\begin{table}
	\caption{Summary of packages used in \hipdemo}
	\label{t:packages}
	\centering
	\begin{tabular}{c p{5cm}}
		\toprule
		Package & Description \\
		\midrule
		HIP~\cite{Rizoiu2017} & An open-source implementation for modeling online popularity with HIP\footnote{HIP code: \url{https://github.com/andrei-rizoiu/hip-popularity/}}.\\
		Shiny \cite{shiny} & An integrated package for implementing web apps in the R language. \\
		ShinyJs \cite{shinyjs} & A tool that allows JavaScript functions to be called inside R programs. \\
		ShinyDashboard \cite{shinydashboard} & Interface which bootstraps and provides several themes for web page UI. \\
		Plotly \cite{plotly} & An interactive plotting library for many programming languages including R.\\
		Shiny Server \cite{shinyserver} & Server for Shiny applications. \\
		youtube-insight \cite{wu2017beyond} & An integrated Youtube data crawler\footnote{youtube-insight code: \url{https://github.com/computationalmedia/youtube-insight/}}. \\
		\bottomrule
	\end{tabular}
\end{table}

%!TEX root = hip-demo.tex

\section{Related work}

Recent results in popularity modeling feature a range of theoretical models to explain and predict online content popularity~\cite{bakshy2011everyone, Martin2016, Mishra2016,Zhao2015,Rizoiu2017}. 
However, the choice of publicly available platforms that implement these models for the non-scientific user is rather limited.

\citet{khosla2014makes} implement a web-based demonstration\footnote{Demonstration from \citet{khosla2014makes}: \url{http://popularity.csail.mit.edu/}} for predicting image popularity before publishing the images. 
Their system computes a popularity confidence by taking image content and a range of social factors into account. 
\citet{castillo2014characterizing} create FAST\footnote{Platform by \citet{castillo2014characterizing}: \url{http://fast.qcri.org/}} (Forecast and Analytics of Social Media and Traffic) for predicting views of news article, by leveraging anonymous real-time page view data. 
\citet{xie2016topicsketch} provide a tool\footnote{Tool from \citet{xie2016topicsketch}: \url{http://research.pinnacle.smu.edu.sg/clear/}} that uses real-time Twitter data to identify potential viral topics and predict their total popularity.

\hipdemo differs from the aforementioned works in several ways. 
First, it aims to be more than a demonstration of a scientific algorithm: it is a visualization platform dedicated to users. 
It has multiple visualization components and the user-friendly interaction enables this platform to easily convey the modeling outcome. 
Second, this platform is built for Youtube video popularity modeling, while other platforms generally deal with Twitter, news article, etc. 
Third, it is open-source and it allows developers to integrate it with other popularity models and additional data sources.

% Popularity models from recent studies can be generally divided into two main categories: \textit{feature-driven} approaches and \textit{generative} approaches. \textit{Feature-driven} approaches apply machine learning algorithms with some preselected features \cite{bakshy2011everyone, Martin2016}, while typical \textit{generative} approaches involve a specific stochastic process model and parameters that are fitted with observed event data and are more interpretable \cite{Mishra2016,Zhao2015,Rizoiu2017}. Every single activity that contributes to popularity can be modeled as a event, after which we are able to treat event data as stochastic processes. Given different types of input data, \textit{generative} approaches further fall into two classes: volume-based models and event-based models. \textit{Event-based approaches} model on individual events and \textit{volume-based approaches} deal with the scenario that only volumes of events are available.

\bibliographystyle{ACM-Reference-Format}
\bibliography{biblio}


\begin{thebibliography}{00}

%%% the Web address.

\ifx \showCODEN    \undefined \def \showCODEN     #1{\unskip}     \fi
\ifx \showDOI      \undefined \def \showDOI       #1{#1}\fi
\ifx \showISBNx    \undefined \def \showISBNx     #1{\unskip}     \fi
\ifx \showISBNxiii \undefined \def \showISBNxiii  #1{\unskip}     \fi
\ifx \showISSN     \undefined \def \showISSN      #1{\unskip}     \fi
\ifx \showLCCN     \undefined \def \showLCCN      #1{\unskip}     \fi
\ifx \shownote     \undefined \def \shownote      #1{#1}          \fi
\ifx \showarticletitle \undefined \def \showarticletitle #1{#1}   \fi
\ifx \showURL      \undefined \def \showURL       {\relax}        \fi
% invisible to TeX
\providecommand\bibfield[2]{#2}
\providecommand\bibinfo[2]{#2}
\providecommand\natexlab[1]{#1}
\providecommand\showeprint[2][]{arXiv:#2}

\bibitem[\protect\citeauthoryear{Attali}{Attali}{2016}]%
        {shinyjs}
\bibfield{author}{\bibinfo{person}{Dean Attali}.}
  \bibinfo{year}{2016}\natexlab{}.
\newblock \bibinfo{booktitle}{{\em shinyjs\: Easily Improve the User Experience
  of Your Shiny Apps in Seconds}}.
\newblock
\showURL{%
\url{https://CRAN.R-project.org/package=shinyjs}}
\newblock
\shownote{R package version 0.9.}


\bibitem[\protect\citeauthoryear{Bakshy, Hofman, Mason, and Watts}{Bakshy
  et~al\mbox{.}}{2011}]%
        {bakshy2011everyone}
\bibfield{author}{\bibinfo{person}{Eytan Bakshy}, \bibinfo{person}{Jake~M
  Hofman}, \bibinfo{person}{Winter~A Mason}, {and} \bibinfo{person}{Duncan~J
  Watts}.} \bibinfo{year}{2011}\natexlab{}.
\newblock \showarticletitle{Everyone's an influencer: quantifying influence on
  twitter}. In \bibinfo{booktitle}{{\em WSDM '11}}. \bibinfo{pages}{65--74}.
\newblock


\bibitem[\protect\citeauthoryear{Castillo, El-Haddad, Pfeffer, and
  Stempeck}{Castillo et~al\mbox{.}}{2014}]%
        {castillo2014characterizing}
\bibfield{author}{\bibinfo{person}{Carlos Castillo}, \bibinfo{person}{Mohammed
  El-Haddad}, \bibinfo{person}{J{\"u}rgen Pfeffer}, {and} \bibinfo{person}{Matt
  Stempeck}.} \bibinfo{year}{2014}\natexlab{}.
\newblock \showarticletitle{Characterizing the life cycle of online news
  stories using social media reactions}. In \bibinfo{booktitle}{{\em CSCW
  '14}}. ACM, \bibinfo{pages}{211--223}.
\newblock


\bibitem[\protect\citeauthoryear{Chang}{Chang}{2016}]%
        {shinydashboard}
\bibfield{author}{\bibinfo{person}{Winston Chang}.}
  \bibinfo{year}{2016}\natexlab{}.
\newblock \bibinfo{booktitle}{{\em shinydashboard: Create Dashboards with
  `Shiny'}}.
\newblock
\showURL{%
\url{https://CRAN.R-project.org/package=shinydashboard}}
\newblock
\shownote{R package version 0.5.3.}


\bibitem[\protect\citeauthoryear{Khosla, Das~Sarma, and Hamid}{Khosla
  et~al\mbox{.}}{2014}]%
        {khosla2014makes}
\bibfield{author}{\bibinfo{person}{Aditya Khosla}, \bibinfo{person}{Atish
  Das~Sarma}, {and} \bibinfo{person}{Raffay Hamid}.}
  \bibinfo{year}{2014}\natexlab{}.
\newblock \showarticletitle{What makes an image popular?}. In
  \bibinfo{booktitle}{{\em WWW '14}}. ACM, \bibinfo{pages}{867--876}.
\newblock


\bibitem[\protect\citeauthoryear{Martin, Hofman, Sharma, Anderson, and
  Watts}{Martin et~al\mbox{.}}{2016}]%
        {Martin2016}
\bibfield{author}{\bibinfo{person}{Travis Martin}, \bibinfo{person}{Jake~M.
  Hofman}, \bibinfo{person}{Amit Sharma}, \bibinfo{person}{Ashton Anderson},
  {and} \bibinfo{person}{Duncan~J. Watts}.} \bibinfo{year}{2016}\natexlab{}.
\newblock \showarticletitle{{Exploring Limits to Prediction in Complex Social
  Systems}}. In \bibinfo{booktitle}{{\em WWW '16}}. \bibinfo{pages}{683--694}.
\newblock
\showDOI{%
\url{https://doi.org/10.1145/2872427.2883001}}


\bibitem[\protect\citeauthoryear{Mishra, Rizoiu, and Xie}{Mishra
  et~al\mbox{.}}{2016}]%
        {Mishra2016}
\bibfield{author}{\bibinfo{person}{Swapnil Mishra},
  \bibinfo{person}{Marian-Andrei Rizoiu}, {and} \bibinfo{person}{Lexing Xie}.}
  \bibinfo{year}{2016}\natexlab{}.
\newblock \showarticletitle{{Feature Driven and Point Process Approaches for
  Popularity Prediction}}. In \bibinfo{booktitle}{{\em CIKM '16}}.
  \bibinfo{pages}{1069--1078}.
\newblock


\bibitem[\protect\citeauthoryear{{R Core Team}}{{R Core Team}}{2013}]%
        {R}
\bibfield{author}{\bibinfo{person}{{R Core Team}}.}
  \bibinfo{year}{2013}\natexlab{}.
\newblock \bibinfo{booktitle}{{\em R: A Language and Environment for
  Statistical Computing}}.
\newblock R Foundation for Statistical Computing, Vienna, Austria.
\newblock
\showURL{%
\url{http://www.R-project.org/}}


\bibitem[\protect\citeauthoryear{Rizoiu and Xie}{Rizoiu and Xie}{2017}]%
        {Rizoiu2017b}
\bibfield{author}{\bibinfo{person}{Marian-Andrei Rizoiu} {and}
  \bibinfo{person}{Lexing Xie}.} \bibinfo{year}{2017}\natexlab{}.
\newblock \showarticletitle{{Online Popularity under Promotion: Viral
  Potential, Forecasting, and the Economics of Time}}. In
  \bibinfo{booktitle}{{\em ICWSM'17}}. \bibinfo{pages}{182--191}.
\newblock


\bibitem[\protect\citeauthoryear{Rizoiu, Xie, Sanner, Cebrian, Yu, and {Van
  Hentenryck}}{Rizoiu et~al\mbox{.}}{2017}]%
        {Rizoiu2017}
\bibfield{author}{\bibinfo{person}{Marian-Andrei Rizoiu},
  \bibinfo{person}{Lexing Xie}, \bibinfo{person}{Scott Sanner},
  \bibinfo{person}{Manuel Cebrian}, \bibinfo{person}{Honglin Yu}, {and}
  \bibinfo{person}{Pascal {Van Hentenryck}}.} \bibinfo{year}{2017}\natexlab{}.
\newblock \showarticletitle{{Expecting to be HIP: Hawkes Intensity Processes
  for Social Media Popularity}}. In \bibinfo{booktitle}{{\em WWW '17}}.
  \bibinfo{publisher}{ACM Press}, \bibinfo{address}{Perth, Australia.},
  \bibinfo{pages}{735--744}.
\newblock
\showISBNx{9781450349130}


\bibitem[\protect\citeauthoryear{{RStudio, Inc}}{{RStudio, Inc}}{2013a}]%
        {shiny}
\bibfield{author}{\bibinfo{person}{{RStudio, Inc}}.}
  \bibinfo{year}{2013}\natexlab{a}.
\newblock \bibinfo{booktitle}{{\em Easy web applications in R.}}
\newblock
\newblock
\shownote{\url{http://www.rstudio.com/shiny/}.}


\bibitem[\protect\citeauthoryear{{RStudio, Inc}}{{RStudio, Inc}}{2013b}]%
        {shinyserver}
\bibfield{author}{\bibinfo{person}{{RStudio, Inc}}.}
  \bibinfo{year}{2013}\natexlab{b}.
\newblock \bibinfo{booktitle}{{\em Shiny Server}}.
\newblock
\newblock
\shownote{URL: \url{https://www.rstudio.com/products/shiny/shiny-server/}.}


\bibitem[\protect\citeauthoryear{Sievert, Parmer, Hocking, Chamberlain, Ram,
  Corvellec, and Despouy}{Sievert et~al\mbox{.}}{2016}]%
        {plotly}
\bibfield{author}{\bibinfo{person}{Carson Sievert}, \bibinfo{person}{Chris
  Parmer}, \bibinfo{person}{Toby Hocking}, \bibinfo{person}{Scott Chamberlain},
  \bibinfo{person}{Karthik Ram}, \bibinfo{person}{Marianne Corvellec}, {and}
  \bibinfo{person}{Pedro Despouy}.} \bibinfo{year}{2016}\natexlab{}.
\newblock \bibinfo{booktitle}{{\em plotly\: Create Interactive Web Graphics via
  `plotly.js'}}.
\newblock
\showURL{%
\url{https://CRAN.R-project.org/package=plotly}}


\bibitem[\protect\citeauthoryear{Wu, Rizoiu, and Xie}{Wu et~al\mbox{.}}{2017}]%
        {wu2017beyond}
\bibfield{author}{\bibinfo{person}{Siqi Wu}, \bibinfo{person}{Marian-Andrei
  Rizoiu}, {and} \bibinfo{person}{Lexing Xie}.}
  \bibinfo{year}{2017}\natexlab{}.
\newblock \showarticletitle{Beyond Views: Measuring and Predicting Engagement
  on YouTube Videos}.
\newblock \bibinfo{journal}{{\em arXiv:1709.02541\/}} (\bibinfo{year}{2017}).
\newblock


\bibitem[\protect\citeauthoryear{Xie, Zhu, Jiang, Lim, and Wang}{Xie
  et~al\mbox{.}}{2016}]%
        {xie2016topicsketch}
\bibfield{author}{\bibinfo{person}{Wei Xie}, \bibinfo{person}{Feida Zhu},
  \bibinfo{person}{Jing Jiang}, \bibinfo{person}{Ee-Peng Lim}, {and}
  \bibinfo{person}{Ke Wang}.} \bibinfo{year}{2016}\natexlab{}.
\newblock \showarticletitle{Topicsketch: Real-time bursty topic detection from
  twitter}.
\newblock \bibinfo{journal}{{\em TKDE\/}} \bibinfo{volume}{28},
  \bibinfo{number}{8} (\bibinfo{year}{2016}), \bibinfo{pages}{2216--2229}.
\newblock


\bibitem[\protect\citeauthoryear{Zhao, Erdogdu, He, Rajaraman, and
  Leskovec}{Zhao et~al\mbox{.}}{2015}]%
        {Zhao2015}
\bibfield{author}{\bibinfo{person}{Qingyuan Zhao}, \bibinfo{person}{Murat~A
  Erdogdu}, \bibinfo{person}{Hera~Y He}, \bibinfo{person}{Anand Rajaraman},
  {and} \bibinfo{person}{Jure Leskovec}.} \bibinfo{year}{2015}\natexlab{}.
\newblock \showarticletitle{{SEISMIC: A Self-Exciting Point Process Model for
  Predicting Tweet Popularity}}. In \bibinfo{booktitle}{{\em ACM SIGKDD
  Conference on KDD}}.
\newblock
\showISBNx{9781450336642}


\end{thebibliography}

\end{document}